\documentclass[12pt]{article}
\usepackage{latexsym}

\begin{document}

\title{ Description of Black Hole Microstates by Means of a Free Affine-Scalar Field}
\author{A. Giacomini \thanks{e-mail: giacomin@science.unitn.it}\\
Dipartimento di Fisica, Universit\`a di Trento}
\date{}
\maketitle

\begin{abstract}
In this article we will invetsigate the origin of central charges in  the Poisson algebra of charges, which arise 
in dimensionally reduced theories describing black holes.  In order to do this we will analyze the equations of motion
arising from the dimensionally reduced Einstein-Hilbert action with the ansatz of spherical symmetry. We will show
that the  equations of motion and the constraints, which involve two fields, namely the dilaton (i.e. the radial coordinate)
and the conformal factor of the 2-metric can be integrated by means of a free field.
The transformation properties of this field must be found. The solution of the free field, which gives 
the Rindler space-time, which is the near-horizon approximation of the Schwarzschild solution, is equivalent to the
solution of the conformal factor of the 2-metric. The conformal factor is a piece of a metric and therefore under conformal
transformations it transforms like an affine scalar. The free field being in this case equivalent to the conformal factor
must therefore also behave like an affine scalar. The stress-energy tensor generating such an affine transformation
is the improved stress-energy tensor. The second derivative term in this tensor gives origin to the central term in the
charges algebra. It is therefore the affine transformation property of the field that gives origin to the central charge
used to compute the black hole entropy via Cardy formula.     

\end{abstract}

\section{Introduction}
Great efforts have been done in order to find a microscopic interpretation of black hole entropy given by the
Bekenstein-Hawking formula \cite{hawking&bardeen&carter}
\cite{hawking} \cite{bekenstein} \cite{bekenstein2}. Such a microscopic interpretation 
should be given by a quantum theory of gravity. The problem is that up to now there does not exist a complete quantum theory of gravity.
There are therefore two different  approaches to the problem. One approach is to use one of the 
the several specific models like superstrings \cite{vafa}, loop quantum gravity \cite{ashtekar} or Sakharov 
induced gravity \cite{sakharov},  to
count the black hole microstates for some classes of black holes. This approaches give always the same Bekenstein-Hawking
result, also being conceptually very different.\\
The second approach is to try to count the black hole microstates in a way that is independent from a specific model.
An interesting approach in this direction is to try to find a 2-dimensional conformal field theory that describes 
the black hole already at classical level. If the charges algebra of such a theory admits a nonzero classical
central charge the black hole microstates can be counted by means of the Cardy \cite{cardy} formula.
This computation is then completely independent from the specific model of quantum theory of gravity.\\
There are different ways of how such a classical central charge can arise.
The first approach is the one done by Carlip  et al.
\cite{carlip}, \cite{ghosh}, \cite{carlip2}, \cite{silva} , where the central charge arises because of the presence
of boundary terms in the canonical generators needed to make them differentiable. This approach seems to have some 
technical problems \cite{park} \cite{soloviev} especially for the Schwarzschild case.
It was then also shown by A. Giacomini \cite{giacomini5} that the Poisson algebra for diffeomorphism generators
that preserve the metric of the black hole bifurcation metric or its surface gravity has zero central charge.
This approach seems to work therefore only  in the case of the BTZ \cite{strominger}
black hole but in this case the asymptotic structure of $AdS_3$ at 
infinity is used instead of the near horizon structure of the black hole becoming a specific case of the $AdS/CFT$ 
conjecture \cite{maldacena}.\\
There is also another way of how to find a classical central charge i.e. studying the dimensionally reduced
Einstein-Hilbert action with spherical symmetry. One obtains a theory with 2 degrees of freedom namely
the dilaton field and the 2-metric of the $r-t$ plane.One can therefore
focus on the dilaton field like in \cite{solodukhin}, \cite{brustein}
and and see if in a near horizon approximation it becomes a 2-D CFT.
A generalization to Gauss Bonnet gravity and higher curvature Lagrangians can be found in \cite{pallua1} \cite{pallua2}. 
The other possibility is, as done by A.Giacomini
and N. Pinamonti \cite{giacomini&pinamonti}
and A. Giacomini \cite{giacomini2} \cite{giacomini3} , to focus on the 2-metric, which can be written in conformally flat form.
\begin{equation}
g_{ij} ^{(2)}= \exp (-2\rho ) \eta _{ij}
\end{equation}
In this case all the geometry of the $r-t$ plane of the black hole is encoded in the conformal factor $\rho$.
In the last cited articles it was shown that in the near horizon approximation the dilaton field is constant
and therefore the only equation of motion that survives is the one for the conformal factor,
which becomes a Liouville theory.  The charges Poisson  algebra of the Liouville theory has a classical
nonzero central charge. It is therefore possible to compute the black hole entropy via Cardy formula.\\
The aim of this article is to understand the origin of such a classical central charge.
In fact we will see that the relevant point for such entropy computations, which use the existence of a
classical central charge, is not the existence of a Liouville theory i.e. a field theory with an exponential
potential, which describes the near horizon black hole. The relevant point is that this field does not transform as 
a scalar field, under Weyl transformations,
\begin{equation}
g_{ij} \rightarrow e^{\omega}g_{ij} \; ,
\end{equation}
but as an affine scalar field
\begin{equation}
\gamma\rho \rightarrow \gamma \rho -\omega  \; , \label{affine}
\end{equation}
where $\gamma$ is a constant.
In the case of the conformal factor $\rho$ it is obvious that this field  cannot be a scalar field
being a piece of a metric. It must therefore transform as an affine scalar. The same situation we have 
in Liouville theory. In fact one obtains the Liouville equation from the equation of constant negative curvature 
surfaces 
\begin{equation}
R[g]= -\lambda
\end{equation}
by putting the metric $g$ in conformally flat form and using the conformal transformation properties of
the curvature scalar. The Liouville field is therefore also an affine scalar field under conformal transformations.
It will be shown that it is this transformation property that is the origin of the central charge in the Poisson algebra,
and not the exponential potential. Indeed we will see that the black hole microstates can be described 
by a free affine-scalar field.\\
In order to do this we will show in the first section that the equations of motion and constraints 
of the dimensionally reduced gravitational action with spherical symmetry, which imply two 
fields i.e. the dilaton and the conformal factor, can be integrated in terms of a free field. 
We want to investigate the near horizon physics of the black hole, therefore in the second section we will analyze the
the form of the free field that produces  the Rindler metric. We will see that this field is equivalent up to a
rescaling to the conformal factor of the Rindler metric, written in conformal flat gauge. This means that the 
free field must transform as an affine scalar field under conformal transformations.
The stress-energy tensor which generates such field transformations is not the usual one of the free field but 
the so called improved energy stress tensor it is the second derivative term in this tensor, that is responsible for the
appearing of a central term in the Poisson algebra of the charges..
This improved stress-energy tensor has in fact in light coordinates 
the same form as the Liouville  stress-energy tensor. Therefore the central charge and the black hole entropy can be 
computed easily using the results of previous articles of the author. The difference to the previous articles is that we don't
need  Liouville theory (i.e. an affine scalar theory with exponential potential) in order to compute the 
black hole entropy. In fact only the affine transformation property is relevant for the computation and therefore it 
possible to describe the near horizon microstates by means of a free field.
In this situation it becomes also clear why the two different approaches to consider or the fluctuations of the
dilaton field or the fluctuations of the conformal factor give always the correct Bekenstein Hawking result.
In fact both fields can be integrated by means of the free field.

\section{Dimensional reduction and integration of the equations of motion}
Let us begin with the usual Einstein-Hilbert action 
\begin{equation}
I= \frac{1}{16\pi}\int \sqrt{-g} R d^4 x  \; . 
\end{equation}
Making the ansatz of spherical symmetry for the static black hole
\begin{equation}
ds^2= g^{(2)}_{ij}dx^i dx^j + \Phi^2 (x_1 ,  x_2) \left( d\theta^2  + \sin ^2 \theta d\phi ^2  \right) \; ,        \label{spheric1}
\end{equation}
and performing the dimensional reduction we obtain the effective action
\begin{equation}
I=\frac{1}{4} \int  d^2 x \sqrt{-g^{(2)}}\left( 2(\nabla \Phi)^2 +\Phi ^2 R^{(2)} +2  \right)  \; .       \label{reduced1}
\end{equation}
We can redefine the 2-metric and the dilaton field $\Phi$ 
\begin{equation}
\Phi^2 = \eta \; \; \; ;  \; \; g_{ab} ^{(2)} = \frac{1}{\sqrt{\eta}} \tilde{g}_{ab}     \label{redefine1}
\end{equation}
obtaining so the action of a dilatonic two dimensional theory in the usual form

\begin{equation}
I=\frac{1}{2} \int \sqrt{-\tilde{g}} \left[ \frac{\eta}{2} R[\tilde{g}] + V(\eta )   \right] \; ,  \label{reduced2}
\end{equation}
where the dilatonic potential is given by
\begin{equation}
V(\eta ) = \frac{1}{\sqrt{\eta}} \; .
\end{equation} 
We will see that the explicit form of the potential is not relevant
for the integration of the equations of motion.
In two dimension we can write the metric in conformally flat form
\begin{equation}
\tilde{g} _{ab} = e^{-2\rho} \gamma _{ab} \; ,                     \label{flatgauge}
\end{equation}
where $\gamma_{ab}$ is as usual the Minkowski metric.
The action (\ref{reduced2}) becomes now
\begin{equation}
I= \frac{1}{2} \int d^2 x \left(\frac{1}{2}\eta R[\gamma ] -\partial _a \eta \partial ^a \rho + V(\eta ) e^{-2\rho}   \right)  \; .     \label{reduced3}
\end{equation}
This is a theory of two fields propagating in flat spacetime.
Notice that the piece containing the curvature scalar $R$ is zero in the
flat gauge case and therefore does not contribute to the equations
of motion. But it gives a contribution to the constraints, which does not
vanish also if we choose then the flat background.
Having fixed the gauge in fact the equations of motion must be implemented
by the constraints
\begin{equation}
\frac{\delta I}{\delta g^{ab}} =T_{ab}=0 \; ,      \label{constraint1}
\end{equation}  
where the variation in made before choosing the flat background.
In light coordinates the equations of motion and the constraint
become
\begin{equation}
\partial _{+} \partial _{-} \rho - \frac{\partial_{\eta} V(\eta )}{4} e^{-2\rho}=0                         \label{motion5}
\end{equation}
\begin{equation} 
\partial _+ \partial _- \eta + \frac{V(\eta )}{2} e^{-2\rho} = 0                                  \label{motion6}
\end{equation}
and for the two constraints $T_{\pm \pm} =T_{11} + T_{22} \pm T_{12} =0$
\begin{equation}
T_{\pm \pm } =\partial _{\pm} \partial _{\pm} \eta + 2 \partial _{\pm} \rho \partial _{\pm} \eta =0 \; .    \label{constraint3}
\end{equation}
Therefore in light coordinates the constraints do not depend on the dilatonic potential.\\
The constraints can be easily integrated
\begin{equation} 
\partial_{\pm} \eta = \exp \left( -2\rho + C_{\mp}(x_{\mp})  \right) \; ,    \label{arbitrary1}      
\end{equation} 
where the function $C_{\mp} $ is an arbitrary function of $x^\mp$. \\
Let us introduce a function $\phi _{\pm} (x^{\pm} ) $ of $x^{\pm} $ defined as 
\begin{equation}
\partial _{\pm} \phi _{\pm} \equiv \exp{(-C_{\pm})}  \; .   \label{phi}
\end{equation} 
So the eq. (\ref{arbitrary1}) becomes 
\begin{equation}
\partial _{\pm} \eta = e^{-2\rho} \, e^{+C_{\mp}} = e^{-2\rho} \left( e^{-C_{\mp}}  \right) ^{-1} = e^{-2\rho} \left(\partial _{\mp} \phi _{\mp}  \right) ^{-1}
\end{equation}
and therefore we can write 
\begin{equation}
e^{-2\rho} -\partial _{\pm} \eta \partial _{\mp} \phi _{\mp} =0 \; .       \label{integration1}
\end{equation}
Inserting this in the second equation of motion(\ref{motion6}) we obtain
\begin{equation}
\partial _+ \partial _- \eta + \frac{V(\eta)}{2} \partial _{\pm} \eta \partial _{\mp} \phi _{\mp} =0  \; .       \label{integration2}
\end{equation}
Let us now introduce a function of the field $\eta$ $F(\eta )$ defined as 
\begin{equation}
\partial _{\eta}  F(\eta )  \equiv V(\eta)      \label{defeta} 
\end{equation}
Using then the fact that $\phi_{\pm}$ is function only of $x^{\pm}$ we can write (\ref{integration2}) as
\begin{equation}
\partial _+ \partial _- \eta + \partial _{\mp} \left( \frac{ F(\eta ) \partial _{\pm} \phi }{2}  \right) =0 \; . 
\end{equation}
Integrating now with respect to $x^{\mp}$ we obtain
\begin{equation}
\partial _{\pm} \eta + \frac{F(\eta ) \partial _{\pm} \phi _{\pm}}{2} = I_{\pm}    \; ,    \label{integration3}
\end{equation}
where $I_{\pm} $ is a function of $x^{\pm}$ . This function is not completely arbitrary. Let us determinate the form of $I_{\pm}$.
In order to do this consider  (\ref{integration2}), and let us write down explicitly the 2 equations it contains  
\begin{equation}
\begin{array}{c}
\frac{2}{V(\eta )}\partial _+ \partial _- \eta = - \partial _+ \eta \partial _- \phi _- \\
\\
\frac{2}{V(\eta ) } \partial _+ \partial _- \eta = - \partial _- \eta \partial _+ \phi _+ \; .
\end{array}
\end{equation}  
We have then 
\begin{equation}
\partial _+ \phi _+ \partial _- \eta = \partial _- \phi _- \partial _+ \eta  \; .   \label{integration4}       
\end{equation}
Let us now take again (\ref{integration3}) which contains also 2 equations 
\begin{equation}
\begin{array}{c}
\partial _+ \eta + \frac{F(\eta )}{2} \partial _+ \phi _+ = I_+ \\
\\
\partial _- \eta + \frac{F(\eta )}{2}\partial _- \phi _- = I_-
\end{array}
\end{equation}
Let us multiply the first one with $\partial _- \phi _- $ and the second one with with $ \partial _+ \phi _+ $ obtaining
\begin{equation}
\begin{array}{c}
\partial _- \phi _- \partial _+ \eta + \frac{F(\eta )}{2} \partial _+ \phi _+ \partial _- \phi _- =I_+ \partial _- \phi _- \\
\\
\partial _+ \phi _+ \partial _- \eta + \frac{F(\eta )}{2}\partial _- \phi _- \partial _+ \phi _+ =I_- \partial _+ \phi _+
\end{array}
\end{equation}
Using now (\ref{integration4}) we obtain the equation
\begin{equation}
I_+ \partial _- \phi _- = I_- \partial _+ \phi _+  \; .            
\end{equation}
In order to satisfy the last equation the function $\phi$ must have the form 
\begin{equation}
I_{\pm} = C_1 \partial _{\pm} \phi _{\pm}  \; ,        \label{integration5}
\end{equation} 
where $C_1$ is a arbitrary constant.
We have so eventually determined the form of $I_{\pm}$. Using the definition of $F(\eta )$ (\ref{defeta}), we notice that $F(\eta )$ is defined up to an
additive constant. We can therefore redefine $F(\eta )$ as 
\begin{equation}
F_{C_1} = F(\eta ) -2C_1    \; ,      \label{newdefeta}
\end{equation}
where the constant $C_1$ is the one introduced in (\ref{integration5}). Using the redefinition  (\ref{newdefeta}) and (\ref{integration5}) we can eliminate
the function $I_{\pm}$ in (\ref{integration3}) which becomes 
\begin{equation}
\begin{array}{c}
\partial _+ \eta = - \frac{F_{C_1}}{2} \partial _+ \phi _+\\
\\
\partial _- \eta = - \frac{F_{C_1}}{2} \partial _- \phi _-
\end{array}            \label{eliminatei}
\end{equation}
which can be integrated 
\begin{equation}
\begin{array}{c}
2\int \frac{\partial _+ \eta }{F_{C_1} (\eta)}dx^+ = -\phi _+ \\
\\
2\int \frac{ \partial _- \eta}{F_{C_1} (\eta )} dx^- =- \phi _- \; .
\end{array}                              \label{integration6}
\end{equation}
Let us now define the field $\phi$ as
\begin{equation}
\phi \equiv \phi _+ + \phi _-   \; .           \label{defphi}
\end{equation} 
Using the fact that
\begin{equation}
d\eta (x^+ , x^- ) = \partial _+ \eta dx^+ + \partial _- \eta dx^-
\end{equation}
equations (\ref{integration6}) can be written as
\begin{equation}
-\phi = 2\int \frac{d\eta}{F_{C_1}(\eta )} +C_2  \equiv 2 G_{C_1 , C_2}  \; ,       \label{integration7} 
\end{equation}
where $C_2$ is an integration constant. The function $\phi$ as defined in   (\ref{defphi}) satisfies obviously the free field equation
\begin{equation}
\Box \phi =0      \label{freefield}  \; .  
\end{equation}
We have expressed the dilaton field $\eta$ in function of a free field.
We can also express the Liouville field $\rho $ in function of $\phi$. Using in fact (\ref{integration1}) and (\ref{eliminatei}) we obtain
\begin{equation}
e^{-2\rho} = \partial _{\pm}\phi \partial _{\mp} \eta = -\frac{F_{C_1}(\eta )}{2} \partial _{\pm} \phi \partial _{\mp} \phi
\end{equation} 
Using (\ref{integration4}) we see that we need to keep only one of the two equations e.g.
\begin{equation}
e^{-2\rho} =- \frac{F_{C_1}(\eta ) }{2}\partial _+\phi \partial _-\phi   \; .  \label{integration8}
\end{equation} 
We started from the equations of motion and the constraints derived from the action (\ref{reduced3}) without any near horizon approximation,
which involved 2 fields namely $\eta$ and $\rho$. We have shown that this two fields can be expressed by means of one free field.
Notice that a free field in two dimensions (and only in two) is conformal field theory. 
Summarizing we obtained
\begin{equation}
\begin{array}{c}
\Box \phi =0  \\
\\
-\phi = 2 G_{C_1 , C_2} (\eta)\\
\\
e^{-2\rho} = - \frac{F_{C_1}(\eta ) }{2} \partial _+ \phi \partial _- \phi
\end{array}                 \label{freefield2}
\end{equation}
Therefore in order to find a conformal field theory with a classical central charge, which can be used to count 
the microstates of the black hole, instead of focosing on the Dilaton field or the conformal factor as in previous papers
we can focus directly on the free field theory. For a scalar free field the charges algebra does not admit a central extension.
The things change if we are dealing with an affine-scalar field as we will see in the next section.  
\section{Near horizon approximation and transformation property of the free field}
Up to now we have found that the reduced equations of motion can be integrated by means of a free field. But we have 
said nothing about the transformation properties of this field. Looking at (\ref{freefield2}) we see that the conformal
factor is function of the free field. The conformal factor is not a scalar and therefore also the free field cannot be a scalar.\\
Let us now make a near horizon approximation.
Near horizon the Schwarzschild $r-t$ plane metric, expanding the lapse function as
\begin{equation}
N^2 (r) =2\kappa (r-r_0) + \mathcal{O}\left(\left( (r-r_0\right)) ^2\right) \; ,
\end{equation}
can be approximated by the Rindler metric
\begin{equation}
ds^2 = -\kappa ^2 y^2 dt^2 +dy^2 \; ,
\end{equation}
which in conformally flat   gauge takes the form
\begin{equation}
ds^2 = -dx^+ dx^- \exp (x^+ - x^- ) \label{rindler} \; .
\end{equation}
Here we have introduced the usual tortoise coordinates $x^{\pm}= \kappa t \pm \log y  $.
Observing the integrated constraints (\ref{arbitrary1}) we see immediately that in the near horizon limit the dilaton
and therefore also the dilatonic potential can be considered constant. Let us therefore study the equations (\ref{freefield2})
in the case of constant dilatonic potential.
Taking
\begin{equation}
V(\eta ) = \lambda  = \mathrm{const} \;.
\end{equation}
Therefore from (\ref{defeta}) we get
\begin{equation}
F(\eta ) = \lambda \eta \; .
\end{equation}
Using (\ref{freefield2}) 
\begin{equation}
G(\eta ) = \frac{1}{\lambda} \log (\lambda \eta) \; .
\end{equation}
Let us now take a free field solution for $\phi$. A possible choice is e.g.
\begin{equation}
\phi = ax^+ - bx^- \;,   \label{sol}
\end{equation}
with constant $a$ and $b$. Using this choice from the second equation of (\ref{freefield2}) we get
\begin{equation}
\frac{2}{\lambda} \log (\lambda \eta ) = - \phi = -(ax^+ - bx^- )
\end{equation}
and so we get for $\eta$
\begin{equation}
\eta = \frac{1}{\lambda} \exp \left( - \frac{\lambda}{2} [ax^+ - bx^- ] \right)  \; .
\end{equation}
Taking now the third equation of (\ref{freefield2}) we obtain
\begin{equation}
\exp (-2\rho ) = \frac{ \lambda \, \eta}{2} \, a b = \frac{ab}{2} \exp \left( \frac{\lambda}{2}[ax^+ - bx^- ] \right) \; ,  
\end{equation}
Taking now for example $a=b=\sqrt{2}$ we have
\begin{equation}
\exp (-2\rho ) = \exp \left( \frac{\lambda \sqrt{2}}{2}[x^+ - x^- ] \right) =\exp ( \frac{\lambda}{2}\phi )
\end{equation}
This choice of $\phi$ therefore gives as solution for the conformal factor the Rindler metric (up to a rescaling).     
We see therefore that up to rescaling the free field (\ref{sol}) is equivalent to the conformal factor $\rho$ of the metric
and therefore transforms in the same way under conformal transformations, namely as an affine scalar (\ref{affine}).\\
Now the stress energy tensor of such an affine-scalar field is different from the stress-energy tensor of a scalar field.
Notice in fact that the usual free field action in two dimensions 
\begin{equation}
I_{sc} = C \int d^2 x \gamma ^{\mu \nu} \partial _{\mu} \phi \partial _{\nu} \phi
\end{equation}
is Weyl invariant if the field $\phi$ transforms as a scalar.\\
Let us analyze this type of action
\begin{equation}
I_{as} = C \int \frac{1}{2} g^{\mu \nu} \nabla _{\mu} \phi \nabla _{\nu} \phi +  R[g]\phi  \; , \label{affineaction}
\end{equation}
where the background metric $g$ is arbitrary. In our case we have chosen the flat background metric, so that the
$R\phi$ coupling does not change the equations of motion. What changes is the form of the stress energy tensor
\[
T_{ab} = \frac{\delta I_{as} }{\delta g_{ab}}
\] 
because the variation of the $R\phi$ term must be taken before choosing the flat background.
In literature this tensor is also called improved stress-energy tensor.
In light coordinates $x^{\pm} = x^1 \pm x^2$ it takes the identical form of the Liouville stress tensor namely
\begin{equation}
T_{\pm \pm} = C\left( (\partial_{\pm} \phi )^2 + \partial _{\pm} ^{2} \phi    \right)   \label{energy}
\end{equation} 
It differs from the usual $T_{\pm \pm}$ of the free field by the last term which has a second derivative. 
It is easy to check that if we smear this tensor with a function $f$ i.e.  $Q_f = \int T_{\pm \pm} f $ then the Poisson 
bracket $\{Q_f , \phi \} $ generates an affine transformation of $\phi$ \cite{jackiv}. \\
Therefore the action $I_{as} $ that describes correctly the free affine scalar field once one chooses the flat background.
This action in fact is Weyl invariant only if the field transforms in an affine way (\ref{affine}).
This affine transformation property reflects also in  the transformation property of the stress energy tensor
and this is the origin of the classical central charge in the Poisson algebra of the charges.

\section{Virasoro algebra and microstates counting}
We can now proceed to construct the Virasoro algebra of the charges. Being the stress tensor (\ref{energy} ) identical to
the Liouville one the computation is identical as in \cite{giacomini&pinamonti} and will be therefore only briefly sketched.
We must first of all determinate the constant $C$ in front of the action (\ref{affineaction}). Being this an effective model
describing a black hole we set the constant in such a way that the energy of the system equals the  mass of the
black hole $M_{BH}$.

\begin{equation}
\int_{-l/ \sqrt{2}} ^{l/ \sqrt{2}} T_{11} dx^2 =M_B \Rightarrow C= \frac{\kappa A }{2\pi l} \; .         \label{costant1}
\end{equation}
The parameter $l$ is a cutoff parameter, that eventually must tend to infinity.
Now we can define the generators
\begin{equation}
L_n ^{\pm} = \int _{-l/ 2\kappa} ^{l/ 2\kappa} \kappa dx^{\pm} \xi_n ^{\pm} T_{\pm \pm}  \; ,         \label{viragen1}
\end{equation}
where the factor $2\kappa$ in the integration is used to match the Euclidean periodicity. 
The $\xi_n$ are smearing fields defined as
 
\begin{equation}
\xi_n ^{\pm} = \frac{l}{2\kappa \pi}\exp \left( -i \frac{2\pi \kappa}{l} n x^{\pm}  \right)   \; .           \label{xi1}
\end{equation}
We can now compute easily the Poisson brackets of the generators
\begin{equation}
\left\{ L_n ^{\pm} , L_m ^{\pm}  \right\} _{PB} =i(n-m)L_{m+n} ^{\pm} + i \frac{c}{12}n^3 \delta _{m+n ,0} \; ,       \label{Virasoro}
\end{equation}
with 
\begin{equation} 
L_0 ^+ = \frac{A l}{16\pi ^2 }        \; \; \; \; \;    c= \frac{3A}{2l}       \;      \label{charge1}
\end{equation}
where having introduced dimensional charges we have redefined the Poisson
brackets as $\{ \dots\} \rightarrow \frac{2\pi l}{A\kappa } \{ \dots \}$. 
Now for every fixed value of the cutoff $l$ we are able to compute the entropy with the logarithm of the  Cardy formula
\begin{equation}
S= 2\pi \sqrt{\frac{c^+ L_0^+}{6} } = \frac{A}{4}  \; ,        \label{bekenstein1}
\end{equation}
which is exactly the Bekenstein-Hawking entropy. The product 
$L_0 c$ in the Cardy formula does not depend on the cutoff parameter
and therefore we can take the limit $l \rightarrow \infty$ also
if the $L_0$ generator diverges.

\section{Conclusions}
We have analyzed the equations of motions and constraints, which arise from the gravitational action with spherical symmetry.
The two fields involved namely the dilaton and the conformal factor can be integrated in terms of a free field.
We concluded therefore that in order to compute the black hole entropy by means of a classical central charge via Cardy formula,
instead of focusing on the dilaton or the conformal factor we can analyze directly this free field.\\
In order to do this we studied the transformation properties of this field. In the near horizon approximation we found that
the free field must transform like an affine scalar. The stress energy tensor which generates such transformations is the improved stress
energy tensor. In light coordinates this is  identical to the stress tensor of the Liouville theory. Having found that, it is easy 
to find the central charge and the black hole entropy using the result of previous papers where near horizon the equation of
motion of the conformal factor became a Liouville equation. The new point in this article is that we found that the relevant fact
for finding the central charge is the affine transformation property of the field and not the existence of an exponential potential.\\
Therefore in order to find a quantum theory of the black hole microstates one can start from this free-affine scalar field
instead of the Liouville field, which will be object of further investigations in future. 

\section*{Acknowledgements}
The author wants to thank Dr. Nicola Pinamonti and Dr. Valter Moretti for many suggestions and interesting discussions on this
subject.

\end{document}